# A Spherical Harmonic Approach to Redshift Distortion and a Measurement of $\Omega_o$ from the 1.2 Jy *IRAS* Redshift Survey

Karl B. Fisher, Caleb A. Scharf and Ofer Lahav
*Institute of Astronomy, Madingley Rd., Cambridge, CB3 0HA, England.*

**ABSTRACT**
We examine the nature of galaxy clustering in redshift space using a method based on an expansion of the galaxian density field in Spherical Harmonics and linear theory. Our approach provides a compact and self-consistent expression for the distortion when applied to flux limited redshift surveys. The amplitude of the distortion is controlled by the combination of the density and bias parameters, $\beta \equiv \Omega_o^{0.6}/b$; we exploit this fact to derive a maximum likelihood estimator for $\beta$. We check our formalism using $N$-body simulations and demonstrate it provides an unbiased estimate of $\beta$ when the amplitude and shape of the galaxy power spectrum is known. Application of the technique to the 1.2 Jy *IRAS* redshift survey yields $\beta \approx 1.0$; both random errors (from counting statistics and the uncertainties in the power spectrum normalization) and systematic errors (from the uncertainty in the shape of the power spectrum) individually contribute 20% uncertainties in this estimate. This estimate of $\beta$ is comparable (both in amplitude and uncertainty) with previous measurements based on comparisons of the *IRAS* density field with direct measurements of peculiar velocities and analyses of the acceleration of the Local Group, but the Spherical Harmonic Analysis has the advantage of being easy to implement and is largely free of systematic errors.

## 1 INTRODUCTION

The clustering of galaxies in redshift space appears systematically different from the clustering that one would observe in real space. As Kaiser (1987) has pointed out, structures tangential to an observer's line-of-sight will appear enhanced in redshift space due to the coherent motions of galaxies on large scales. Using linear theory, Kaiser was able to deduce a simple expression for the resulting direction dependent amplification of the redshift space density field. Kaiser found that the redshift space power spectrum was *anisotropic* and related to the real space power spectrum (assumed to be isotropic) by

$$P_S(\mathbf{k}) = P_R(k)\left(1 + \beta\mu_{KL}^2\right)^2 , \quad (1)$$

where $\mu_{KL}$ is the cosine of the angle between the wave vector $\mathbf{k}$ and the observer's line-of-sight. In Equation (1) and in the remainder of this paper, we adopt the notation $\beta \equiv \Omega_o^{0.6}/b$, where $\Omega_o$ and $b$ are the current density and bias (assumed here and below to be independent of scale) parameters respectively; we also will use $R$ and $S$ super and subscripts when referring to quantities in real and redshift space respectively. The quantity $\beta$ naturally appears in expressions for the redshift distortion since it is the proportionality factor between the velocity and density field in linear theory, $\nabla \cdot \mathbf{v} = -H_o\beta\delta$ (e.g., Peebles 1980).

While the redshift distortion given by Equation (1) is quite simple in the Fourier domain, it leads to a fairly complicated expression when cast in terms of the redshift space correlation function, $\xi_S(\mathbf{s})$. The full angular dependence of $\xi_S(\mathbf{s})$ was first derived by Lilje & Efstathiou (1989) and then by McGill (1990) and Hamilton (1992) in various forms. We find it illuminating to write $\xi_S(\mathbf{s})$ in the form

$$\xi_S(\mathbf{s}) = \left[1 + \frac{2}{3}\beta + \frac{1}{5}\beta^2\right] A_0(s)\mathcal{P}_0(\mu_{LS})$$
$$- \left[\frac{4}{3}\beta + \frac{4}{7}\beta^2\right] A_2(s)\mathcal{P}_2(\mu_{LS}) + \left[\frac{8}{35}\beta^2\right] A_4(s)\mathcal{P}_4(\mu_{LS}) , \quad (2)$$

where $A_l(s) = \frac{1}{2\pi^2} \int dk\, k^2 P_R(k) j_l(ks)$, $j_l(x)$ is the spherical Bessel function and $\mathcal{P}_l(\mu_{LS})$ is the usual Legendre polynomial evaluated at the cosine of the angle between **s** and the observer's line-of-sight. In principle, the dependence of the redshift distortions on $\beta$ offers the hope that $\Omega_o$ can be determined dynamically (e.g., Hamilton 1992, 1993) assuming the bias can be determined independently.

In practice there are several disadvantages to analysing the redshift distortions with $\xi_S(\mathbf{s})$. Firstly, Kaiser's analysis is valid only in the linear regime where $\delta(\mathbf{r}) \ll 1$. Unfortunately, $\xi_S(\mathbf{s})$ depends on cumulative moments of the real space correlation function which can contain contributions from nonlinear evolution on small scales (although see Hamilton *et al.* 1991 for a discussion of moments which are unaffected by nonlinear evolution). Secondly, Equation (1) is strictly only applicable to volume limited samples. In the case of a flux limited sample characterized by a radial selection function, $\phi(r)$, (the number density in a homogeneous flux limited survey being $\propto r^2 \phi(r)$) there is a contribution to the redshift distortion which is given by

$$\left(2 + \frac{d\ln\phi}{d\ln r}\right) \left[\frac{U(\mathbf{r}) - \mathbf{V}_{obs} \cdot \hat{\mathbf{r}}}{r}\right] \quad , \tag{3}$$

where we have adopted Kaiser's notation for the radial component of the peculiar velocity field, $U(\mathbf{r}) = \mathbf{v}(\mathbf{r}) \cdot \hat{\mathbf{r}}$ and $\mathbf{V}_{obs}$ is the velocity of observer defined in the same reference frame as $\mathbf{v}(\mathbf{r})$. Although Kaiser considered the effect of a radial selection function in his analysis, this term was assumed to be small and was omitted from Equation (1).

In this paper, we present a new and complementary way of investigating redshift distortions based on a linear expansion of the galaxian density field in spherical harmonics. Spherical Harmonic analysis (hereafter SHA) leads to a surprisingly simple expression for the redshift distortion which naturally includes the effect of a radially varying selection function. Moreover, since the $l^{\text{th}}$ order harmonic probes a typical scale $\pi D/l$ (where $D$ is roughly the sample depth), one can ensure the analysis is carried out in the linear regime by restricting the investigation to the low order multipole moments.

Scharf *et al.* (1992), Lahav *et al.* (1993), and Scharf & Lahav (1993; hereafter SL93) have shown that SHA provides a useful method for probing the local cosmography of the galaxy distribution. During their investigation of galaxy clustering using SHA, they noted the sensitivity of their results to assumptions regarding the nature of the galaxy peculiar velocity field. The present work is an extension of the SHA which statistically incorporates the effect of redshift distortion in a self-consistent way.

We begin the paper by outlining the SHA method in § 2. In § 3, we check our formalism by comparing our predictions for the harmonics in redshift space with those actually computed from numerical $N$-body simulations and show how it can be used to formulate a maximum likelihood estimator for $\beta$. In § 4, we apply our formalism to the *IRAS* 1.2 Jy redshift survey and derive an estimate of $\beta$. We conclude in § 5.

## 2  METHOD

Following SL93, we define a weighted spherical harmonic decomposition of the flux-limited density field, in redshift space as,

$$a_{lm}^S = \sum_{i=1}^{N_g} f(\mathbf{s}_i) Y_{lm}(\hat{\mathbf{s}}_i) \quad , \tag{4}$$

where $N_g$ is the number of galaxies in the survey, $Y_{lm}$ is the usual spherical harmonic, and $f(s)$ is an arbitrary radial weighting function.[*] In the special case of constant weighting all redshift distortions vanish and one simply obtains the projected angular distribution on the sky. The harmonics derived from such weighting can, for example, be used to derive the real space correlation function, $\xi(r)$, using techniques developed by Peebles (1973). However, in the more general case where $f(s)$ varies with $s$, peculiar velocities will distort the harmonics.

In order to relate the redshift harmonics given by Equation (4) to their real space counterparts, we first rewrite the summation in Equation (4) as a continuous integral over the density fluctuations in redshift space,

$$a_{lm}^S = \int d^3\mathbf{s}\,\phi(r)f(s)\left[1 + \delta_S(\mathbf{s})\right]Y_{lm}(\hat{\mathbf{s}}) \quad . \tag{5}$$

In Equation (5), the selection function is assumed to be normalized such that $\int dr\, r^2 \phi(r) = N_g/\omega$ where $\omega$ is the solid angle subtended by the survey. Notice that the selection function in Equation (5) is evaluated at the galaxy's *distance*, not redshift, because if the catalogue is flux limited the probability of a galaxy being at redshift, $\mathbf{s}$, will be proportional to the selection function evaluated at the galaxy's actual (albeit unknown) distance, i.e. $\propto \phi(r)$. Next, we note that, by construction, $n_S(\mathbf{s})d^3\mathbf{s} = n_R(\mathbf{r})d^3\mathbf{r}$, where $n_S(\mathbf{s})$ and $n_R(\mathbf{r})$ refer to the densities in redshift and real space and that if the perturbations induced by peculiar motions are small, then we can perform a Taylor series expansion of all redshift quantities to first order in the density fluctuation, e.g., $f(\mathbf{s}) \simeq f(r) + \frac{df(r)}{dr}(U(\mathbf{r}) - \mathbf{V}_{obs} \cdot \hat{\mathbf{r}})$. Thus, expansion of Equation (5) yields,

$$\begin{aligned} a_{lm}^S &= \int d^3\mathbf{r}\,\phi(r)f(r)\left[1 + \delta_R(\mathbf{r})\right]Y_{lm}(\hat{\mathbf{r}}) \\ &+ \int d^3\mathbf{r}\,\phi(r)\frac{df}{dr}(U(\mathbf{r}) - \mathbf{V}_{obs} \cdot \hat{\mathbf{r}})Y_{lm}(\hat{\mathbf{r}}) \quad , \end{aligned} \tag{6}$$

There are several comments to be made regarding the validity of Equation (6). Firstly, we have assumed that the summation in Equation (4) is carried out over *all* galaxies in a flux limited redshift survey. In this case, the integrals in Equation (6) extend over all space and there are no "surface" terms arising from the deformation of the boundary of the integration region that occurs in the transformation from redshift to real space. Secondly, there is an apparent absence of terms involving the derivatives of the selection function; these terms would be manifest if we, like Kaiser (1987), took $f(s)$ to be the special case of $1/\phi(s)$.

The expansion in Equation (6) can be simplified by expanding the radial peculiar velocity field dependence, $U(\mathbf{r})$, in spherical harmonics:

$$\begin{aligned} U(\mathbf{r}) &\equiv \mathbf{v}(\mathbf{r}) \cdot \hat{\mathbf{r}} \\ &= \frac{\beta}{2\pi^2}\sum_{lm}(i^l)^* \int d^3\mathbf{k}\,\frac{\delta_\mathbf{k}^R}{k}j_l'(kr)Y_{lm}(\hat{\mathbf{k}})Y_{lm}^*(\hat{\mathbf{r}}) \quad . \end{aligned} \tag{7}$$

---

[*] In our analysis $f(s)$ is required to be continuous in its first derivative and to vanish at infinity; this simplifies the analysis by eliminating surface terms that arise when $f(s)$ has a discontinuous boundary.

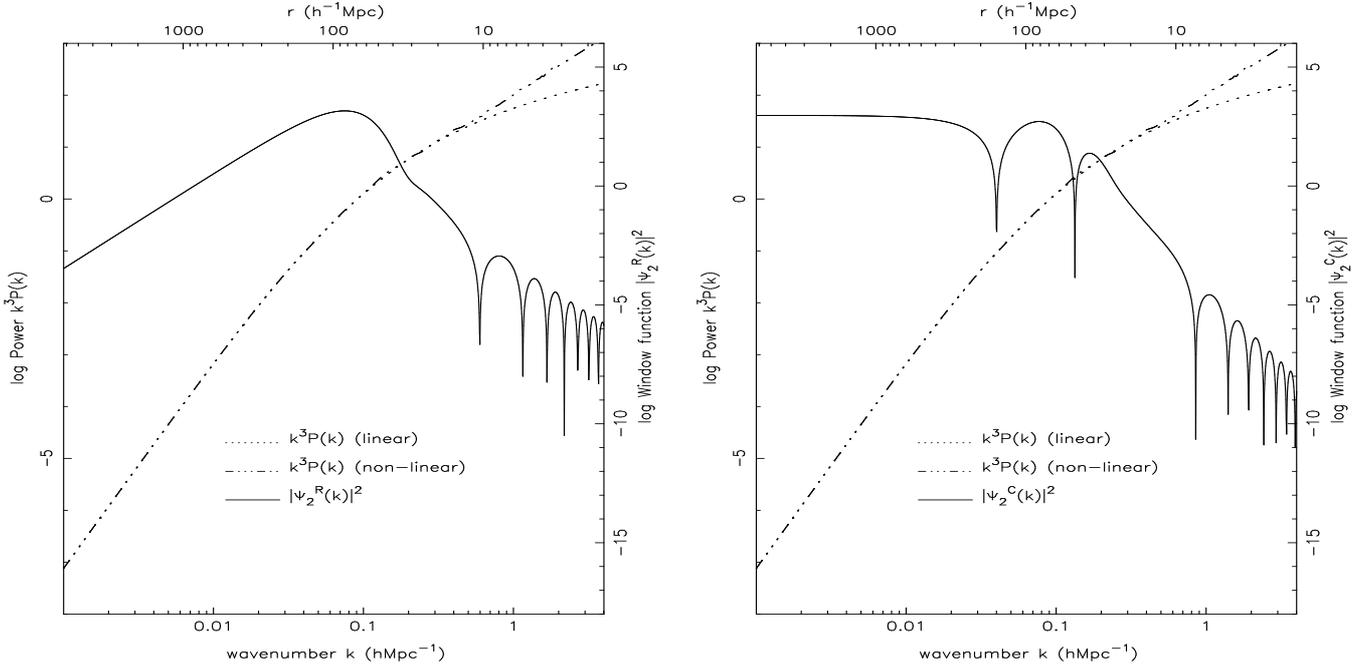

**Figure 1.** An example of the window functions appearing Equations (9) and (10). In each panel, the dotted curves show $k^3 P(k)$ for the CDM ($\Omega_0 h = 0.5$) linear theory power spectrum and the dot-dashed curves show the corresponding CDM nonlinear power spectrum determined from the $N$-body simulation described in the text. In the left panel, the solid curve represents the $|\Psi_2^R(k)|^2$, the window function corresponding the real space contribution to the quadrupole ($l = 2$) moment as defined in Equation (9). The solid curve in the right hand panel respresents the window function of the redshift distortion (again for $l = 2$), $|\Psi_2^C(k)|^2$ defined in Equation (10). In deriving the window functions we took the weighting function, $f(r)$, to be a Gaussian with mean of 40 $h^{-1}$ Mpc and a dispersion of 15 $h^{-1}$ Mpc; this corresponds to the window used construct the harmonics in Figure 2.

In Equation (7) and below, $j_l'(kr) = dj_l(kr)/d(kr)$ refers to the first derivative of the Bessel function. In Equation (7), we have used the continuity equation to relate the velocity to the density field, $\mathbf{v_k} = -i\beta \delta_{\mathbf{k}}^R \mathbf{k}/k^2$; in this limit Equation (6) can be written (for $l \geq 1$),

$$a_{lm}^S = \frac{(i^l)^*}{2\pi^2} \int d^3\mathbf{k}\, \delta_{\mathbf{k}}^R \left[\Psi_l^R(k) + \beta \Psi_l^C(k)\right] Y_{lm}(\hat{\mathbf{k}}), \quad (8)$$

where

$$\Psi_l^R(k) = \int_0^\infty dr\, r^2 \phi(r) f(r) j_l(kr) \quad (9)$$

describes the real space contribution to the harmonics and

$$\Psi_l^C(k) = \frac{1}{k} \int_0^\infty dr\, r^2 \phi(r) \frac{df(r)}{dr} \left(j_l'(kr) - \frac{1}{3}\delta_{l1}\right), \quad (10)$$

is a "correction" term which embodies the redshift distortions. In Equation (10), $\delta_{l1}$ is a Kronecker delta which contributes only to the dipole ($l = 1$) harmonic. In the case of equal weight, $df/dr = 0$, $\Psi_l^C(k)$ vanishes and Equation (8) leads to the real space expression for the harmonics derived by Scharf *et al.* (1992). The expressions given in Equations (8), (9), and (10) are exact in linear theory.

The window functions, $\Psi_l^R(k)$ and $\Psi_l^C(k)$, corresponding to a Gaussian weighting function $f(s)$ with mean 40 $h^{-1}$ Mpc and dispersion 15 $h^{-1}$ Mpc are shown in Figure 1. The distortion term, $\Psi_l^C(k)$, (the right hand panel) is sensitive to the nature of the velocity field and therefore to the long wavelength modes of the density field; in contrast, the real space window, $\Psi_l^R(k)$, is relatively insensitive to large scale fluctuations.

The expected linear theory *rms* value of the harmonics is given by ensemble average of the square of Equation (8). Using the statistical independence of Fourier waves[†], $\langle \delta_{\mathbf{k}}^R \delta_{\mathbf{k}'}^{R*} \rangle = (2\pi)^3 P_R(k) \delta^{(3)}(\mathbf{k} - \mathbf{k}')$, this yields

$$\langle |a_{lm}^S|^2 \rangle = \frac{2}{\pi} \int_0^\infty dk\, k^2 P_R(k) \left|\Psi_l^R(k) + \beta \Psi_l^C(k)\right|^2. \quad (11)$$

In real data the square of harmonics in Equation (11) will have a discreteness or "shot" noise contribution; this can be modeled by adding

$$\langle |a_{lm}|^2 \rangle_{SN} = \int_0^\infty dr\, r^2 \phi(r) [f(r)]^2 \quad (12)$$

to Equation (11).

The $l = 1$ or dipole distortion has two contributions: one from the external dipole moment of the velocity field around the observer and one induced by the motion of observer itself. The latter distortion is caused by $\mathbf{V}_{obs} \neq \mathbf{0}$ and is the origin of the so-called "rocket effect" discussed by Kaiser (1987), Kaiser & Lahav (1988), and Strauss et al. (1992a). We stress, however, that in linear theory the choice of reference frame (i.e., the value of $\mathbf{V}_{obs}$) affects only the value of the dipole harmonic and that for $l \geq 2$ the results are independent of the assumed motion of the observer. The *rms* averaging procedure in Equation (11) treats the observer as a typical point in space and therefore assumes that the observer's peculiar motion is not unusually high or low. This point should be kept in mind if the motion of the Local Group is atypical. In practice, these complications can be avoided by working in the comoving frame of the observer, i.e., the reference frame in which $\mathbf{V}_{obs}$ vanishes.

---

[†] This assumes that the survey contains many independent modes of the wavenumber in question and usually phrased somewhat loosely as the "fair" sample hypothesis.

## 3  $N$-BODY TESTS: A WAY TO DETERMINE $\Omega_\circ$

In order to check the validity of our formalism, we computed both the real and redshift space weighted harmonics in an $N$-body simulation of a standard Cold Dark Matter universe characterized by $\Omega_o h = 0.5$. The simulations evolved $64^3$ particles in a box of $180$ $h^{-1}\,\mathrm{Mpc}$ using the $P^3M$ algorithm until *rms* variance of the density field in a sphere of $8$ $h^{-1}\,\mathrm{Mpc}$ reached $\sigma_8 = 0.61$; further details of the simulations can be found in Frenk et al. (1990). In an effort to mimic current observational data, we extracted nine mock galaxy catalogues designed to closely match the properties of the 1.2 Jy *IRAS* redshift survey (cf., Strauss et al. 1992b, Fisher 1992). Galaxy candidates were chosen as unbiased tracers of the underlying particle distribution and therefore $\beta = 1$ in the mock catalogues. The procedure for extracting these catalogues is described in Górski et al. (1989) and Fisher et al. (1993a).

Figure 2 shows the mean weighted harmonic power spectrum, $C_l^2$, defined as the azimuthally averaged harmonic by the shot noise contribution to the harmonic (cf., Scharf et al. 1992),

$$C_l^2 = \frac{\frac{1}{2l+1}\sum_{m=-l}^{+l}|a_{lm}|^2}{\langle a_l^2\rangle_{SN}} \qquad . \qquad (13)$$

The harmonics were computed both in real and redshift space by performing the direct summation given in Equation (4), squaring the resulting harmonic and then subtracting the shot noise contribution as given in Equation (12).

In deriving the harmonics shown in Figure 2, we used a somewhat ad hoc Gaussian weighting function for $f(s)$ with centroid at $40$ $h^{-1}\,\mathrm{Mpc}$ and dispersion $\sigma = 15$ $h^{-1}\,\mathrm{Mpc}$. The redshifts in the simulation were corrected to the comoving frame of the observer, i.e., the frame where $\mathbf{V}_{obs} \cdot \hat{\mathbf{r}} = 0$. The solid symbols in Figure 2 show the mean value of the real space harmonic spectrum computed from the harmonics in each of the nine mock *IRAS* catalogues. The open symbols correspond to the same calculation performed in redshift space.

The error bars correspond to the variance in the harmonic estimates over the nine simulations. The solid and dashed curves correspond to the theoretical prediction given by Equation (11), using the known real space power spectrum and $\beta$ of the simulations. From Figure 2, we see that for our choice of window function, the redshift space har-monics are enhanced and that the overall amplification is described quite well by the prediction of Equation (11).

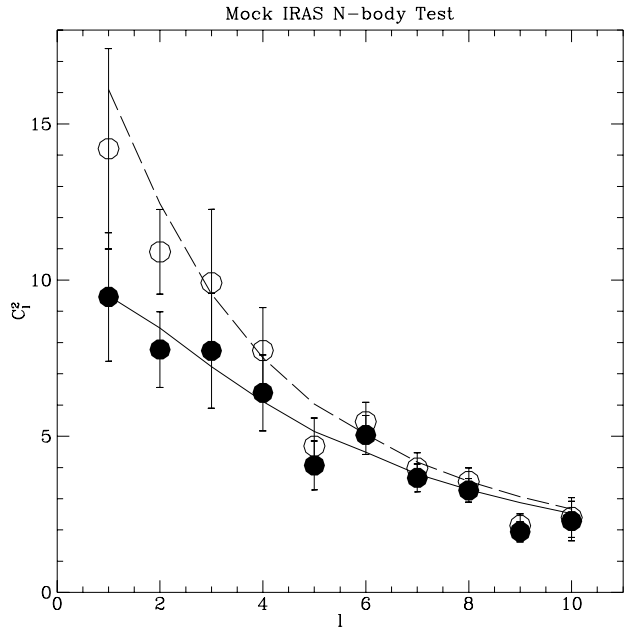

**Figure 2.** The harmonic power spectrum in real space (filled symbols) and redshift space (open symbols) for the mean of nine mock *IRAS* catalogues. The solid and dashed lines represent the linear theory predictions (cf., Equation (11)) for real and redshift space respectively.

Since the distortion is well described by our formalism, the next step is to ask whether one can predict the value of $\beta$ given a set of harmonics measured from a redshift catalogue. The predicted distortion depends on knowing the correct shape and amplitude of the galaxy power spectrum. The normalization of the power spectrum can be derived by looking at the amplitude of the number weighted ($f(s) = 1$) harmonics since in that case the redshift distortion is identically zero; this was the approach adopted by SL93. Alternatively, the amplitude of the power spectrum in real space can be fixed by looking at the real space correlation function. As we will see in the next section, this approach is feasible for redshift surveys large enough to allow the real space correlation function to be deprojected from the redshift space correlation.

The shape of the power spectrum can be conveniently parametrized by a series of phenomenological CDM models with with varying $\Gamma \equiv \Omega_o h$ (e.g., Efstathiou, Bond, & White 1992). Of course, in the $N$-body simulations we know the power spectrum precisely; however in order to apply the formalism to actual data we must quantify the dependence of the derived $\beta$ on the shape of the power spectrum.

The weighting function, $f(r)$, used to construct the harmonics can be chosen arbitrarily. Clearly, one would like to pick the weight function which simultaneously maximizes the distortion while minimizing the noise. We are currently working on this optimization problem (cf., § 5 below). However, the arbitrariness of the weighting function can be somewhat circumvented by computing the harmonics not with a single weighting function, but with a series of different weight functions each of which samples a different region of space. After experimenting with several different choices of weight functions we decided to use four Gaussian windows centered at 38, 58, 78, and 98 $h^{-1}$Mpc each with a dispersion of 8 $h^{-1}$Mpc. For each of the windows, we computed the corresponding weighted redshift harmonics for each of the mock *IRAS* catalogs. Next, for a given choice of power spectrum, we maximized the likelihood given in Equation 14 over a grid in $\beta$. Since our analysis is valid only in the linear regime, we restricted the likelihood computation to $l \leq 10$.

The likelihood of measuring a set of redshift harmonics with $N$ weighting functions, $a_{lm}^i \{i = 1, \ldots N\}$ is given by

$$\ln \mathcal{L} \equiv -\left(\frac{2l+1}{2}\right) \ln |A| - \frac{1}{2} \sum_{i,j} A_{ij}^{-1} \left[Re(a_{l0}^i) Re(a_{l0}^j)\right] \quad (14)$$

$$- \sum_{i,j} A_{ij}^{-1} \sum_{m=1}^{l} \left[Re(a_{lm}^i) Re(a_{lm}^j) + Im(a_{lm}^i) Im(a_{lm}^j)\right] \quad ,$$

where $A_{ij}$ is the covariance matrix

$$A_{ij} = \langle a_{lm}^i a_{lm}^{j*} \rangle_{TH} + \langle a_{lm}^i a_{lm}^{j*} \rangle_{SN} \quad , \quad (15)$$

with

$$\langle a_{lm}^i a_{lm}^{j*} \rangle_{TH} = \frac{2}{\pi} \int_0^\infty dk\, k^2 P_R(k) \left(\Psi_l^{R(i)}(k) + \beta \Psi_l^{C(i)}(k)\right) \times$$

$$\left(\Psi_l^{R(j)}(k) + \beta \Psi_l^{C(j)}(k)\right)^* \quad (16)$$

and

$$\langle a_{lm}^i a_{lm}^{j*} \rangle_{SN} = \int_0^\infty dr\, r^2 \phi(r) \left[f^{(i)}(r) f^{(j)}(r)\right] \quad , \quad (17)$$

corresponding to the different weighting functions, $f^{(i)}(s)$. In Equation 14 $|A|$ is the absolute value of the determinant of $A_{ij}$, and $Re(a_{lm})$ and $Im(a_{lm})$ refer to the real and imaginary parts of $a_{lm}$ (SL93). The superscripts on $\Psi^R(k)$ and $\Psi^C(k)$ refer to the window function of the corresponding weighting function. Since the harmonics at different $l$ are independent, the total likelihood for $l \leq l_{max}$ is just given by the sum of the likelihoods in Equation 14 for each $l$ up to $l_{max}$. For our choice of weight functions, the off-diagonal elements of the covariance matrix, $A_{ij}$, are $\lesssim 10\%$ of the largest diagonal elements; the four spatial windows defined by the weighting functions therefore give roughly independent estimates of $\beta$.

Table 1 shows the resulting estimates of $\beta$ for the nine mock catalogs when the correct normalization and shape of the power spectrum were used in the likelihood. The mean value of $\beta$ for the nine mock samples is $0.94 \pm 0.2$ which is in excellent agreement with the actual value $\beta = 1$. This quantifies the qualitative agreement between the predicted and observed distortion seen in Figure 2; when the correct power spectrum is used the likelihood estimate for $\beta$ is unbiased. The formal uncertainty in a single realization (given by $\Delta \mathcal{L} = -1/2$) is comparable to the scatter over the nine realizations, i.e., $\Delta \beta = 0.2$.

How do the estimates of $\beta$ change when the incorrect power spectrum is used in the likelihood? Figure 3 shows the likelihood as a function of $\beta$ for a single mock *IRAS* catalog when the shape and normalization of the power spectrum are altered from their true values. The left hand panel shows the result of varying the normalization of the true power spectrum by $\pm 10\%$. When the normalization is too low by 10% the corresponding maximum liklehood value of $\beta$ is systematically too high by roughly 40%; the converse is true when too high a normalization is used. In the right hand panel of Figure 3 we show the values of $\beta$ for a power spectrum with more large scale power ($\Gamma = 0.2$) than used in the simulations. This power spectrum predicts excessive power on large scales and therefore the best fit value of $\beta$ is systematically low.

**TABLE 1**
CDM MONTE CARLO RESULTS

| Realization | $\beta$ |
|---|---|
| 1 | 0.93 |
| 2 | 1.11 |
| 3 | 1.06 |
| 4 | 0.86 |
| 5 | 0.83 |
| 6 | 0.88 |
| 7 | 1.15 |
| 8 | 1.13 |
| 9 | 0.51 |
| Mean: 0.94 | |
| Standard Deviation: 0.20 | |

## 4 $\Omega_o$ FROM THE 1.2 JY *IRAS* SURVEY

We now proceed to apply the formalism described in the previous section to redshift survey of 5313 *IRAS* galaxies flux-limited to 1.2 Jy at 60$\mu$m, selected from the *IRAS* database (Strauss et al. 1990; Fisher 1992). The sample covers 87.6% of the sky and is complete for $|b| > 5°$ with the exception of a small area of the sky not surveyed by *IRAS*. The SHA analysis discussed above relies on complete $4\pi$ steradian sky coverage. Although statistical corrections can be applied to the harmonics with partial sky coverage (Scharf et al. 1992; SL93), we have adopted a simpler method of dealing with incomplete sky coverage. We have interpolated the redshift data through the plane in a way which smoothly continues structure; this is the same approach used by Yahil et al. (1991) in their reconstruction of the *IRAS* peculiar velocity field. Moreover, regularized mask inversion of the harmonics shows the effect of incomplete sky coverages is small for the *IRAS* 1.2 Jy sample geometry and $l \leq 10$ (Lahav et al. 1993, in preparation). The selection function, $\phi(r)$, is computed using the techniques of Yahil et al. (1991). In practice the selection function is computed by assigning luminosities based on galaxy redshifts, not distances; fortunately, the shape of the $\phi(r)$ is insensitive to the flow model used to correct for the effects of peculiar velocities (Saunders et al. 1990).

The typical redshift error in the 1.2 Jy *IRAS* survey is $\sim 50$ km s$^{-1}$. We repeated the *N*-body tests discussed in the previous section with the redshifts of the simulation particles perturbed by a Gaussian deviate with dispersion

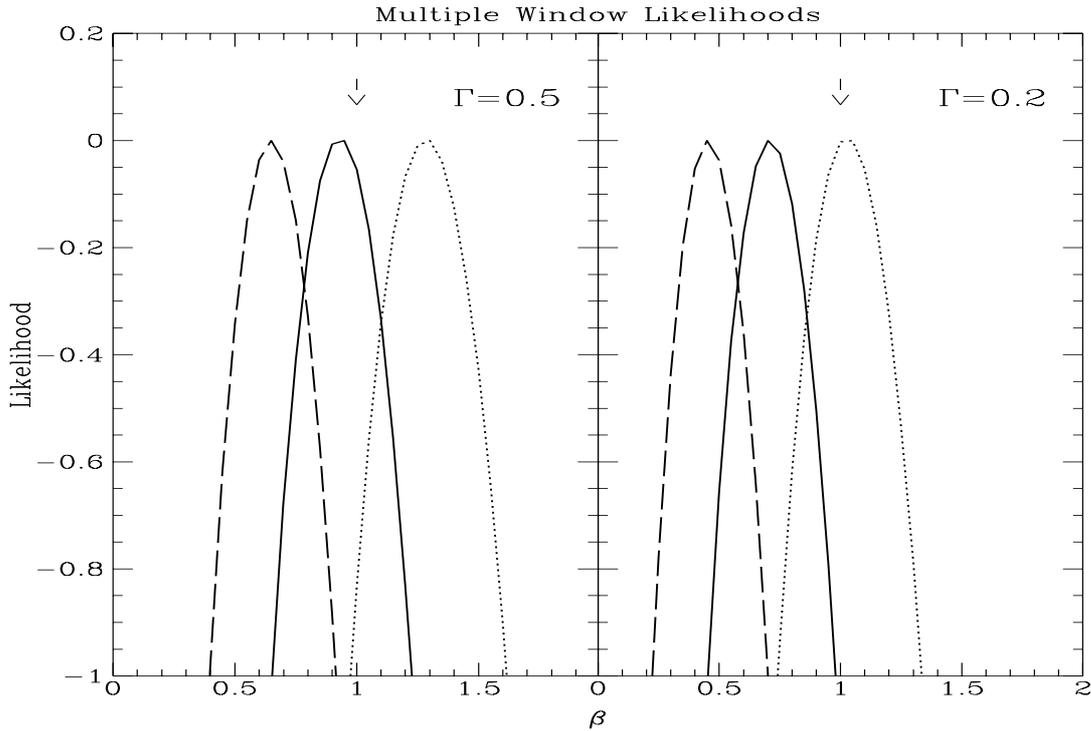

**Figure 3.** Likelihood curves ($\ln \mathcal{L}$) for $\beta$ for a *single* mock *IRAS* realization. The left panel corresponds to likelihood when the correct ($\Gamma = \Omega_o h = 0.5$) power spectrum used to construct the mock catalog is used, while the right panel corresponds to the likelihood if the model predictions are made with a power spectrum with more large scale power ($\Gamma = 0.2$). The solid curve in each panel is the likelihood curve obtained using the value of $\sigma_8$ derived from the projected harmonics ($f(r) =$constant) for the assumed power spectrum. The dotted and dashed curves correspond to the likelihoods when $\sigma_8$ is perturbed down and up (respectively) by 10%. The arrows in each panel indicate the true value of $\beta$ in the simulation, $\beta = 1$.

50 km s$^{-1}$. The resulting value of $\beta$ was stable to within 2%. As a further check we divided the simulations into two halves and then perturbed the redshifts in one half with a Gaussian of dispersion 30 km s$^{-1}$ and the other half with a dispersion of 70 km s$^{-1}$ in an effort to simulate possible systematic errors in the redshifts between the Northern and Southern hemispheres; again the derived $\beta$ was remarkably stable, giving us confidence that the technique is not biased significantly by measurement error in the redshifts.

The correlation function in *real* space has been determined for the 1.2 Jy survey (Fisher *et al.* 1993b). The real space correlation function can be integrated analytically to derive the variance in a 8 $h^{-1}$Mpc sphere $\sigma_8$ which is the conventional normalization of the power spectrum. Fisher *et al.* (1993b) found $\sigma_8 = 0.69 \pm 0.04$. We have adopted this normalization for the power spectrum used in the likelihood. The shape of the *IRAS* power spectrum is somewhat more uncertain but appears consistent with a CDM like model with $\Gamma \approx 0.2$ (Fisher *et al.* 1993a; Feldmann, Kaiser, & Peacock 1993).

In Table 2, we show the maximum likelihood estimates of $\beta$ from the 1.2 Jy survey for $\Gamma = 0.5$ and 0.2 with $\sigma_8$ fixed at 0.69 and when $\sigma_8$ is perturbed up and down by its $1-\sigma$ errors. If *IRAS* is well described by a $\Gamma = 0.2$ spectrum then the corresponding best value of $\beta$ is 0.96; uncertainties in the normalization give a 20% variation in this estimate. The derived value of $\beta$ for a true CDM spectrum ($\Gamma = 0.5$) is higher $\beta \approx 1.3 \pm 0.2$. The errors in the normalization of the power spectrum and the formal uncertainty from the likelihood estimator should be independent. Therefore the random error in $\beta$ is roughly $2^{1/2} 0.2 \sim 0.3$. The systematic uncertainty arising from the shape of the power spectrum is $\lesssim 0.2$.

TABLE 2
IRAS 1.2 Jy $\beta$ ESTIMATES
$\Gamma = \Omega_o h$

|  | 0.50 | 0.20 |
|---|---|---|
| $\sigma_8 = 0.69$ | 1.34 | 0.96 |
| $\sigma_8 = 0.65$ | 1.58 | 1.16 |
| $\sigma_8 = 0.73$ | 1.13 | 0.78 |

## 5 DISCUSSION

The value of $\beta$ derived from the 1.2 Jy sample using the SHA technique, $\beta \approx 1.0 \pm 0.3$, is in agreement with previous estimates from *IRAS* redshift surveys. Dekel *et al.* (1993) found $\beta = 1.28^{+0.75}_{-0.59}$ ($2\sigma$) by comparing the density field of the 1.936 Jy *IRAS* survey (Strauss *et al.* 1992b) with the density field inferred from direct measurements of the peculiar velocity field using the POTENT algorithm (Bertschinger & Dekel 1989; Dekel, Bertschinger, & Faber 1990). Kaiser et al. (1991) found $\beta = 0.9^{+0.20}_{-0.15}$ (1$\sigma$) in a comparison of the predicted *IRAS* peculiar velocity field with direct measurements using the deeper, albeit sparser, QDOT redshift survey. Measurements based on the predicted acceleration of the Local Group using *IRAS* catalogs have also yielded high estimates $0.4 < \beta < 1.0$ (Rowan-Robinson et al. 1990; Strauss et al. 1992a). Hamilton's (1993) analysis of redshift distortions in the correlation function for the 1.936 Jy *IRAS* survey gave a slightly lower value, $\beta = 0.66^{+0.34}_{-0.22}$ (1$\sigma$), although consistent with the SHA to within the quoted 1$\sigma$ error.

These results provide strong evidence of a high, near closure, value of $\Omega_0$ on large scales *if* the bias factor for *IRAS* is near unity. However, one should be cautious in claiming that these results provide definitive proof of $\Omega_0 = 1$. Similar estimates of $\beta$ using optical surveys have yielded lower values ($0.25 < \beta < 0.65$) for the optical dipole (Lynden-Bell, Lahav, & Burstein 1989) and ($\beta \sim 0.5$) in comparisons of optical samples with peculiar velocities (Hudson 1993, in preparation)) suggesting that at least the ratio of the *IRAS* and optical bias factors may differ from unity.

The estimate of $\beta$ obtained with the SHA is largely free of systematic errors. The main systematic error arises from uncertainties in the shape of the *IRAS* power spectrum on large scales; analyses of the extended QDOT redshift survey (Saunders et al. in prepartion) should greatly reduce this uncertainty. The SHA method has the further advantage of requiring only redshift data. This should be constrasted with the POTENT technique which, although conceptually very elegant, can be affected by systematic errors in the direct measurements of peculiar velocities (e.g., the Malmquist bias). Unlike dipole analyses of the Local Group acceleration which are plagued by systematic errors arising from redshift distortions (the "rocket" effect discussed in § 2), the SHA treats the redshift distortion in an internally self-consistent way. Estimates of $\beta$ from the redshift correlation function $\xi_s$ (Hamilton 1993) are prone to systematic errors arising from non-linear distortions on small scales and are extremely noisy on larger scales where the signal to noise ratio in the estimate of the correlation function is low. Morever, linear analyses based on Kaiser's formulation of the distortion neglect the term $\left(2 + \frac{d \ln \phi}{d \ln r}\right)\left[\frac{U(\mathbf{r}) - \mathbf{V}_{obs} \cdot \hat{\mathbf{r}}}{r}\right]$, (Kaiser 1987, equation 3.3); although the relative importance of this term falls with distance, it should be included when analysing the redshift space clustering of local, and usually well sampled, structure.

We are currently examining the question of how to design an optimal weighting scheme that will yield a minimum variance estimate of $\beta$ using techniques similar to those discussed in Strauss et al. (1992a). Several main criteria should be considered when picking the functional form of the different weighting functions. Firstly, the derivative of the window function should be nonvanishing since the distortion indentically vanishes when $df/dr = 0$. Secondly, the various weighting functions should sample different regions of space to avoid a high covariance between the structure sampled. Thirdly, the windows defined by the weighting functions should not give excessive weight to distant galaxies which will be poorly sampled in current redshift surveys. Fourthly, the windows should sample a large enough volume to ensure that the harmonics are indicative of their *rms* values. One should keep in mind, however, that the statistical error for the weighting scheme used in our analysis was comparable to the systematic error induced by uncertainties in the shape of the power spectrum. Consequently, further optimization will improve the quoted errors only if coupled with more reliable estimates of the power spectrum on large scales.

The SHA formalism for the redshift distortion in galaxy samples is also very convenient for comparison with two other major probes of the cosmic structure, peculiar velocities and the microwave background. The difference between the real and redshift space harmonics can be written (cf., Equations (9), (9), and (11)) for $l \geq 2$ as

$$\langle |a^S_{lm} - a^R_{lm}|^2 \rangle =$$

$$\Omega_0^{1.2} \frac{2}{\pi} \int\limits_0^\infty dk\, P_M(k) \left| \int\limits_0^\infty dr\, r^2 \phi(r) \frac{df(r)}{dr} j'_l(kr) \right|^2 , \quad (18)$$

where $P_M(k) = P_R(k)/b^2$ is the power spectrum of the matter fluctuations. ‡

This is entirely analogous to expressions for bulk flows and peculiar velocities. In particular, it is similar to the expansion of the radial velocity field in harmonics (Regös & Szalay (1989), equations 44-45) :

$$\langle |V_{lm}|^2 \rangle = \Omega_0^{1.2} H_0^2 \frac{2}{\pi} \int\limits_0^\infty dk\, P_M(k) \left| \int d^3\mathbf{r}\, \Phi(\mathbf{r}) j'_l(kr) \right|^2 , \quad (19)$$

where $\Phi(\mathbf{s})$ is the generalized (anisotropic) selection function for the observed radial velocities. A comparison of Equations (18) and (19) shows that the redshift distortion seen in the SHA can be directly related to the *rms* multipole moments of the peculiar velocity field.

Equations (9), (10), and (11) are also very similar to the harmonics of the temperature fluctuations in the cosmic microwave background due to the Sachs-Wolfe effect,

$$\langle |a_{lm}|^2 \rangle_{SW} = \left(\frac{H_0}{2c}\right)^4 \frac{2}{\pi} \int\limits_0^\infty \frac{dk}{k^2} P_M(k) [j_l(2ck/H_0)]^2 , \quad (20)$$

for $\Omega_0 = 1$ (cf., Bertschinger 1992). Relations 18 and 19 suggest the possibility of determining, for a given power-spectrum in real-space, the product $\Omega_0^{0.6}\sigma_8$ where $\sigma_8$ is the normalization of the power spectrum of the entire *matter* distribution, independent of biasing.

In summary, the SHA formalism therefore offers a new approach to the study of redshift distortions and the estimation of $\Omega_0$ which is conceptually rich and allows the unification of a variety of cosmic phenomena, from cosmography to quantitative statistical studies, into a single mathematical language.

---

‡ Strictly speaking the bias factor relating the galaxy and mass correlation functions is not the same as the bias that enters into Equations 1 and 11. Only in the constant bias model assumed throughout this paper are the two equal.

**Acknowledgments**

We thank A.J.S. Hamilton, P. Lilje, D. Lynden-Bell and H. Mo for helpful comments and discussions. We also thank M. Strauss, J.P. Huchra, M. Davis, and A. Yahil for permission to use the 1.2 Jy *IRAS* survey data prior to its publication. KBF and CAS acknowledge a SERC postdoctoral fellowship and studentship respectively.

**REFERENCES**

Bertschinger, E. 1992, in New Insights into the Universe, ed. V. J. Martinez, M. Portilla & D. Saez (New York: Springer-Verlag), in press

Bertschinger, E. & Dekel, A. 1989, ApJL, 336, L5

Dekel, A., Bertschinger, E., & Faber, S.M. 1990, ApJ, 364, 349

Dekel, A., Bertschinger, E., Yahil, A., Strauss, M.A., Davis, M., & Huchra, J.P. 1993, ApJ, in press

Efstathiou, G., Bond, R., & White, S.D.M. 1992, MNRAS, 258, 1P

Feldmann, H., Kaiser, N., & Peacock, J. 1993, preprint

Fisher, K. B. 1992, Ph.D. thesis, University of California, Berkeley

Fisher, K. B., Davis, M., Strauss, M. A., Yahil, A., & Huchra, J. P. 1993a, ApJ, 402, 42

Fisher, K. B., Davis, M., Strauss, M. A., Yahil, A., & Huchra, J. P. 1993b, MNRAS, in press

Frenk, C. S., White, S. D. M., Efstathiou, G., & Davis, M. 1990, ApJ, 351, 10

Górski, K., Davis, M., Strauss, M. A., White, S. D. M., & Yahil, A. 1989, ApJ, 344, 1

Hamilton A. J. S. 1992, ApJL, 385, L5

Hamilton A. J. S. 1993, ApJL, 406, L47

Hamilton A. J. S., Kumar, P., Lu, E., & Matthews, A. 1991, ApJL, 374, L1

Kaiser, N. 1987, MNRAS, 227, 1

Kaiser, N., Efstathiou, G., Ellis, R., Frenk, C., Lawrence, A., Rowan-Robinson, M., & Saunders, W. 1991, MNRAS, 252, 1

Kaiser, N. & Lahav, O. 1988, in Large-Scale Motions in the Universe, ed. V. C. Rubin, G. V. Coyne, S. J. (Princeton: Princeton University Press), p. 366

Lahav, O., Yamada, T., Scharf, C., & Kraan-Korteweg, R. C. 1993, MNRAS, 262, 711

Lilje, P. & Efstathiou, G. 1989, MNRAS, 236, 851

Lynden-Bell, D., Lahav, O., & Burstein, D. 1989, MNRAS, 234, 677

McGill, C. 1990, MNRAS, 242, 428

Regős, E. & Szalay, A. S. 1989, ApJ, 345, 627

Rowan-Robinson, M. *et al.* 1990, MNRAS, 247, 1

Peebles, P.J.E. 1980, The Large-Scale Structure of the Universe, (Princeton: Princeton University Press)

Peebles, P.J.E. 1973, ApJ, 185, 413

Saunders, W., Rowan-Robinson, M., Lawrence, A., Efstathiou, G., Kaiser, N., Ellis, R.S., & Frenk, C.S. 1990, MNRAS, 242, 318

Scharf, C., Hoffman, Y., Lahav, O., & Lynden-Bell, D. 1992, MNRAS, 256, 229

Scharf, C. & Lahav, O. 1993, MNRAS, in press

Strauss, M.A., Davis, M., Yahil, A., & Huchra, J.P. 1990, ApJ, 361, 49

Strauss, M. A., Huchra, J. P., Davis, M., Yahil, A., Fisher, K. B. & Tonry, J. 1992b, ApJS, 83, 29

Strauss, M. A., Yahil, A., Davis, M., Huchra, J. P., & Fisher, K. B. 1992a, ApJ, 397, 395

Yahil, A., Strauss, M.A., Davis, M., & Huchra, J.P. 1991, ApJ, 372, 380